\documentclass[prd,twocolumn]{revtex4-1}

\def\mnras{MNRAS}

\begin{document}

\title{Real space {CMB} temperature and polarization formulas}

\author{M. Landriau}\email{mlandriau@lbl.gov}
\affiliation{McDonald Observatory,
2515 Speedway, Stop C1402, Austin, Texas 78712, USA}
\affiliation{Lawrence Berkeley National Laboratory, 1 Cyclotron Road,
  Mailstop 50R5008, Berkeley, California 94720, USA}

\begin{abstract}
I derive formulas for the real space CMB temperature and
polarization by solving the first order Boltzmann equation for the
Stokes parameters $I$, $Q$ and $U$.
\end{abstract}

\maketitle

\section{Introduction}

Most studies of the CMB that use all the relevant physics to first
order in the cosmological perturbations,
i.e. by solving the first order Einstein-Boltzmann equations,
limit themselves to working exclusively in harmonic space, in order to
directly compute the power spectra,
e.g.~\cite{seljak96}.
In real space, we have the Sachs-Wolfe formula for CMB temperature anisotropies, which was derived
assuming instantaneous recombination~\cite{sachs67}.  Although there
is an \textit{ad hoc} way
to take the finite duration of recombination into account,
e.g.~\cite{pen94,methods}, the
resulting formula is not exact, as it's missing a term, which has been
known for some time (at least for scalar modes)
from harmonic space computations, e.g. the seminal work of
\cite{bond87}.
Studying maps of CMB fluctuations is
desirable when one is interested in looking for localized features,
which are not readily identifiable in power spectra.
In~\cite{hrcmb}, the full real space temperature
equation was used to compute CMB temperature maps seeded by a network
of cosmic strings.  Here, I will
present the proof of this formula as well as derive those for polarization.

\section{The Boltzmann equation}

We want to find real solutions to the first order Boltzmann equation for the Stokes
parameters $I$, $Q$ and $U$, which, following \cite{melchiorri96} we
normalize in the following manner:
\begin{equation}
\left(\begin{array}{c}I\\Q\\U\end{array}\right)=\frac{\rho_{\gamma} (\eta)}{4\pi}
\left(\begin{array}{c}1+\Delta_I\\ \Delta_Q\\ \Delta_U
\end{array}\right)\, ,
\end{equation}
where $\rho_{\gamma}$ is the photon energy density and $\eta$ is the
conformal time.  The equations are then
\begin{equation}\begin{array}{c}
\frac{\partial\Delta_I}{\partial\eta} + \hat{n}\cdot\nabla\Delta_I 
+2 \hat{n}_i \hat{n}_j \frac{\partial h_{ij}}{\partial\eta}
= \dot{\tau}(\tilde{\Delta}_I - \Delta_I +4\hat{n}\cdot
\mathbf{v}_b)\, ,
\\
\frac{\partial\Delta_{Q,U}}{\partial\eta}  +
\hat{n}\cdot\nabla\Delta_{Q,U} = \dot{\tau}(\tilde{\Delta}_{Q,U}
- \Delta_{Q,U})\, ,
\end{array}\end{equation}
where $\hat{n}$ is the direction of photon propagation,
$\dot{\tau}$ is the differential Thompson cross section,
$\mathbf{v}_b$ and $h_{ij}$ are the baryon velocity and
metric perturbations in the synchronous gauge and
\begin{equation}\label{eq:scat_def}
\left(\begin{array}{c}\tilde{\Delta}_I(\hat{n})\\
    \tilde{\Delta}_Q(\hat{n})\\ \tilde{\Delta}_U(\hat{n})\end{array}\right)
= \int\mathcal{M} (\hat{n},
  \hat{n}^{\prime})\left(\begin{array}{c}\Delta_I(\hat{n}^{\prime})\\
      \Delta_Q(\hat{n}^{\prime})\\ \Delta_U(\hat{n}^{\prime})\end{array}\right)
d\Omega^{\prime}
\end{equation}
are the scattering terms.
For the detailed form of the scattering matrix $\mathcal{M}$, the
reader is referred to~\cite{melchiorri96} or \cite{methods}; its
derivation can be found in the classic monograph \cite{chandrasekhar50}.
Going to Fourier space, the equations become
\begin{equation}\label{eq:boltzmann}\begin{array}{c}
\dot{\Delta}_I + \dot{\imath}k\mu\Delta_I + 2\dot{h}_{ij}\hat{n}_i
\hat{n}_j = \dot{\tau} (\tilde{\Delta}_I -
(\Delta_I-4\hat{n}\cdot\mathbf{v}_b))\, ,\\
\dot{\Delta}_{Q,U} + \dot{\imath}k\mu\Delta_{Q,U}  =
\dot{\tau}(\tilde{\Delta}_{Q,U}-\Delta_{Q,U})\, ,
\end{array}\end{equation}
where $\mu=\cos\theta=\hat{n}\cdot\hat{k}$.  These can then be formally integrated to yield:
\begin{equation}\label{eq:soln1}\begin{array}{ll}
\Delta_I = &\displaystyle\int d\eta
e^{-ik\mu(\eta_0-\eta)}e^{-\tau}\times\\
&\left(\dot{\tau}
    (\tilde{\Delta}_I+4\hat{n}\cdot\mathbf{v}_b) -
    2\dot{h}_{ij}\hat{n}_i\hat{n}_j\right)\, , \\
\Delta_{Q,U} = &\displaystyle\int d\eta
e^{-ik\mu(\eta_0-\eta)}e^{-\tau}\dot{\tau}
\tilde{\Delta}_{Q,U}\, .
\end{array}\end{equation}
To evaluate the scattering term, we split the $\Delta_{I,Q,U}$ into
scalar, vector and tensor components, and factor out the dependence on
the azimuthal angle $\varphi$; which is
defined with respect to an arbitrary
basis for the plane normal to $\hat{k}$, $\hat{e}_1$ and $\hat{e}_2$
that obey $\hat{e}_1 \times \hat{e}_2 = \hat{k}$, such that
$\hat{e}_1\cdot\hat{n}= \sin\theta\cos{\varphi}$ and 
$\hat{e}_2\cdot\hat{n}= \sin\theta\sin{\varphi}$:
\begin{equation}\begin{array}{ll}
\Delta_I = &\Delta_I^S -\dot{\imath}(1-\mu^2)^{\frac12}(\Delta_{I}^{V1}\cos{\varphi}+
\Delta_{I}^{V2}\sin{\varphi})\\
& + (1-\mu^2)(\Delta_{I}^{T+}\cos{2\varphi}+
\Delta_{I}^{T\times}\sin{2\varphi})\, ,\\
& \\
\Delta_Q = &\Delta_Q^S + \mu (1-\mu^2)^{\frac12}(\Delta_{Q}^{V1}\cos{\varphi}+
\Delta_{Q}^{V2}\sin{\varphi})\\
& + (1+\mu^2)(\Delta_{Q}^{T+}\cos{2\varphi}+
\Delta_{Q}^{T\times}\sin{2\varphi})\, ,\\
& \\
\Delta_U = &(1-\mu^2)^{\frac12}(-\Delta_{U}^{V1}\sin{\varphi}
+\Delta_{U}^{V2}\cos{\varphi})\\
& + 2\mu(-\Delta_{U}^{T+}\sin{2\varphi}
+\Delta_{U}^{T\times}\cos{2\varphi}) \, .
\end{array}
\end{equation}
We then expand the components into Legendre polynomials  $\Delta(\mu)
= \sum_{\ell}(-\dot{\imath})^{\ell} (2\ell
+1)\Delta_{\ell}P_{\ell}(\mu)$.
Performing the integral (\ref{eq:scat_def}), we obtain
\begin{equation}\label{eq:scat_sol2}\begin{array}{ll}
\tilde{\Delta}_I = &\Delta_{I0}^S - \frac14(3\mu^2-1)\Pi^S\\
& + \mu(1-\mu^2)^{\frac12}
\left(\Pi^{V1}\cos\varphi+\Pi^{V2}\sin\varphi\right)\\
&  -(1-\mu^2)
\left(\Pi^{T+}\cos{2\varphi} + \Pi^{T\times}\sin{2\varphi}\right)\, ,\\
& \\
\tilde{\Delta}_Q = & \frac32(\mu^2-1)\Pi^S \\
& \displaystyle+\mu(1-\mu^2)^{\frac12}(\Pi^{V1}\cos\varphi +\Pi^{V2}\sin\varphi)\\
& + (1+\mu^2)(\Pi^{T+}\cos{2\varphi} +
  \Pi^{T\times}\sin{2\varphi})\, ,\\
& \\
\tilde{\Delta}_U = &\displaystyle(1-\mu^2)^{\frac12}(\Pi^{V1}\sin\varphi - \Pi^{V2}\cos\varphi)\\
& + (2\mu)(\Pi^{T+}\sin{2\varphi} - \Pi^{T\times}\cos{2\varphi})\, ,
\end{array}
\end{equation}
with the components $\Pi^{S,V,T}$ defined as in Ref. \cite{methods}:
\begin{equation}\label{eq:PiDef}\begin{array}{ll}
\Pi^{S}=& \Delta^{S}_{I2}+\Delta^{S}_{Q0}+\Delta^{S}_{Q2}\, ,\\
& \\
\Pi^{Vi}=& -\frac{3}{10}\Delta^{Vi}_{I1} -\frac{3}{10}\Delta^{Vi}_{I3}\\
& +\frac{7}{20}\Delta^{Vi}_{Q0}
+\frac{5}{28}\Delta^{Vi}_{Q2}-\frac{6}{35}\Delta^{Vi}_{Q4}\, ,\\
&\\
\Pi^{T\epsilon}=& \frac{1}{10}\Delta^{T\epsilon}_{I0}+
\frac{1}{7}\Delta^{T\epsilon}_{I2}+\frac{3}{70}\Delta^{T\epsilon}_{I4}-\\
&\frac{3}{5}\Delta^{T\epsilon}_{Q0}+\frac{6}{7}\Delta^{T\epsilon}_{Q2}-
\frac{3}{70}\Delta^{T\epsilon}_{Q4}\, .
\end{array}\end{equation}
By introducing the traceless tensor
\begin{equation}\label{eq:poltensdef}
\begin{array}{ll}
\Pi_{ij} = &\frac{3}{4}\Pi^S (\hat{k}_i\hat{k}_j - \frac13\delta_{ij}) \\
& -\frac{1}{2}\Pi^{V1} (\hat{k}_i\hat{e}_{1j} +
  \hat{e}_{1i}\hat{k}_j) - \frac12\Pi^{V2} (\hat{k}_i\hat{e}_{2j} + 
  \hat{e}_{2i}\hat{k}_j) \\
& +\Pi^{T+}(\hat{e}_{1i}\hat{e}_{2j} +
\hat{e}_{2i}\hat{e}_{1j}) +
\Pi^{T\times}(\hat{e}_{1i}\hat{e}_{1j} -
\hat{e}_{2i}\hat{e}_{2j})\, ,
\end{array}\end{equation}
which I call the polarization tensor
\footnote{Note that the prefactors for $\Pi^S$ and the
  $\Pi^V$s suggest that they should have been defined
differently in~\cite{methods}, but for consistency, the same definition was
kept.}
, and the vectors
\begin{equation}
\hat{u} = -\frac{\hat{k} -
  \mu\hat{n}}{\sin\theta} \, ,
\,\,\,\,\,\,\,\,\,\,
\hat{v} = 
\frac{\hat{k}\times\hat{n}}{\sin\theta} \, ,
\end{equation}
which form an orthonormal basis for the polarization plane (i.e. normal to
$\hat{n}$), such that $\hat{u}$ lies in the $\hat{n} - \hat{k}$ plane,
we can simplify the equations for the scattering terms:
\begin{equation}\label{eq:scat_sol}
\left(\begin{array}{c}\tilde{\Delta}_I - \Delta_{I0}^S\\
    \tilde{\Delta}_Q\\ \tilde{\Delta}_U\end{array}\right)
= \Pi_{ij} \left(\begin{array}{c}\hat{n}_i\hat{n}_j\\
    \hat{u}_i\hat{u}_j - \hat{v}_i\hat{v}_j \\
    \hat{u}_i\hat{v}_j + \hat{v}_i\hat{u}_j \end{array}\right) \, .
\end{equation}
Note that the triad $\hat{n}$, $\hat{u}$ and $\hat{v}$ corresponds to
the standard spherical coordinate unit vectors $\hat{r}$,
$\hat{\theta}$ and $\hat{\varphi}$ with $\hat{k}$ as the $z$-axis.
Having determined the form of the scattering terms, we can now derive
the real space formulas for the Stokes
parameters.

\section{CMB temperature and polarization}

From the intensity part of Eqs. (\ref{eq:soln1}) and
(\ref{eq:scat_sol}), the CMB temperature formula can immediately be
obtained by Fourier transforming back to real space and using
$\Delta T/T = \Delta_I/4$ and $\delta_{\gamma} = \Delta_{I0}$:
\begin{equation}\label{eq:temp}
\begin{array}{ll}
\displaystyle\frac{\Delta T}{T}(\hat{n})
=
\displaystyle\int d\eta e^{-\tau} &
\left[
\dot{\tau}\left(\frac{\delta_\gamma}4
-\hat{n}\cdot {\mathbf v}_b
+\frac14\hat{n}_i\hat{n}_j\Pi_{ij}\right)\right.\\
& \left. - \frac12\hat{n}_i\hat{n}_j \dot{h}_{ij}\right] \, .
\end{array}
\end{equation}
This equation for the temperature fluctuations
is essentially the Sachs-Wolfe formula
modified to take into account the finite duration of decoupling with
an added polarization term, $\frac14\hat{n}_i\hat{n}_j\Pi_{ij}$.
Note that we have changed the sign of $\hat{n}$, to change the
perspective from photon propagation direction to line of sight.

For $Q$ and $U$, it is not as simple as, unlike $I$, they are not
invariant under rotations; they are transformed as
\begin{equation}\begin{array}{rcl}
\Delta_Q' &=& \Delta_Q\cos(2\vartheta) + \Delta_U\sin(2\vartheta)\, ,\\
\Delta_U' &=& -\Delta_Q\sin(2\vartheta) + \Delta_U\cos(2\vartheta)\, ,
\end{array}\end{equation}
when rotated by $\vartheta$ about $\hat{n}$.  Since at each point in Fourier
space, the coordinate system defined by $\theta$ and
$\varphi$ (or equivalently $\hat{u}$ and $\hat{v}$) has a different
orientation, we need to rotate by $\xi$ about
$\hat{n}$ (see Fig.~\ref{fig:angles}), in order for the contribution
from each Fourier mode to $Q$ and $U$ to have the same orientation in the
laboratory frame before being integrated over.  Using
the laws of sines and cosines for spherical triangles, we obtain the following
relations for $\xi$:
\begin{equation}\label{eq:xidef}
\begin{array}{c}
\displaystyle\frac{\sin\xi}{\sin\theta_k} = \frac{\sin(\varphi_k - \varphi_n)}{\sin\theta} =
\frac{\sin\varphi}{\sin\theta_n} \, , \\
\\
\cos{\theta_k} = \mu\cos{\theta_n}+ \sin{\theta}\sin{\theta_n}\cos{\xi} \, ,
\end{array}
\end{equation}
where $\theta_k$ and $\varphi_k$ are the polar and azimuthal angles
that define $\hat{k}$ in spherical coordinates in the laboratory frame
and, similarly, $\theta_n$ and $\varphi_n$ are the polar and azimuthal angles
that define $\hat{n}$.
\begin{figure}
\begin{picture}(190,190)(-70,-70)
\thicklines
\put(0,0){\line(0,1){100}}
\put(0,105){\makebox{$z$}}
\put(0,0){\line(1,0){100}}
\put(105,0){\makebox{$y$}}
\put(0,0){\line(-1,-1){50}}
\put(-56,-56){\makebox{$x$}}
\thinlines
\put(0,0){\vector(3,4){66}}
\put(69,92){\makebox{$\hat{k}$}}
\put(0,0){\vector(1,2){40}}
\put(43,73){\makebox{$\hat{n}$}}
\put(66,88){\line(0,-1){110}}
\put(0,0){\line(3,-1){66}}
\put(40,80){\line(0,-1){110}}
\put(0,0){\line(4,-3){40}}
\qbezier(0,80)(20,85)(40,80)
\put(13,73){\makebox{$\theta_n$}}
\qbezier(-7,-7)(7,-15)(21,-7)
\put(6,-16){\makebox{$\varphi_k$}}
\qbezier(0,80)(33,99)(66,88)
\put(33,96){\makebox{$\theta_k$}}
\qbezier(-21,-21)(3,-27)(28,-21)
\put(0,-30){\makebox{$\varphi_n$}}
\qbezier(35,81)(40,87)(45,81)
\put(45,84){\makebox{$\xi$}}
\qbezier(20,40)(27,50)(33,44)
\put(28,48){\makebox{$\theta$}}
\qbezier(40,80)(60,84)(66,88)
\end{picture}
\caption{Angle definitions.}\label{fig:angles}
\end{figure}
This allows us to write the equation for $\Delta_Q$ as
\begin{equation}\begin{array}{ll}
\Delta_Q = & 
\displaystyle\int d\eta\, e^{-ik\mu(\eta_0-\eta)}e^{-\tau}\dot{\tau}
 \,\Pi_{ij}\times\\
& \left((\hat{u}_{i}\hat{u}_{j} -
\hat{v}_{i}\hat{v}_{j}) \cos2{\xi} + 
  (\hat{u}_{i}\hat{v}_{j} +
\hat{v}_{i}\hat{u}_{j}) \sin{2\xi}\right) \, ,
\end{array}
\end{equation}
with the equation for $\Delta_U$ being the same except $\cos{2\xi}$ is
replaced by $-\sin{2\xi}$ and $\sin{2\xi}$ by $\cos{2\xi}$.
It follows from above that the angle of rotation between the bases
$(\hat{u},\, \hat{v})$ and $(\hat{\theta}\, , \hat{\varphi})$, the
latter defined in the usual way
\begin{equation}\begin{array}{c}
\hat{\theta} =
(\cos{\theta_n}\cos{\varphi_n}, \,  \cos{\theta_n}\sin{\varphi_n}, \, -
\sin{\theta_n})\, ,\\
\hat{\varphi} = (-\sin{\varphi_n}, \, 
\cos{\varphi_n}, 0)\, ,
\end{array}\end{equation}
 is $\xi$.  This can also be
verified by taking the dot products of the two bases and using the
equations (\ref{eq:xidef}). Explicitly, we then have
\begin{equation}\begin{array}{c}
\hat{u} = \hat{\theta}\cos\xi - \hat{\varphi}\sin\xi\, , \\
\hat{v} = \hat{\theta}\sin\xi + \hat{\varphi}\cos\xi\, ,
\end{array}\end{equation}
from which we obtain:
\begin{equation}
\begin{array}{ll}
\hat{\theta}_i\hat{\theta}_j - \hat{\varphi}_i\hat{\varphi}_j = & (\hat{u}_{i}\hat{u}_{j} -
\hat{v}_{i}\hat{v}_{j}) \cos2{\xi}\\
 & + (\hat{u}_{i}\hat{v}_{j} +
\hat{v}_{i}\hat{u}_{j}) \sin{2\xi} \, ,\\
& \\
\hat{\theta}_i\hat{\varphi}_j +
\hat{\varphi}_i\hat{\theta}_j
 = &-(\hat{u}_{i}\hat{u}_{j} -
\hat{v}_{i}\hat{v}_{j}) \sin2{\xi} \\
 & + (\hat{u}_{i}\hat{v}_{j} +
\hat{v}_{i}\hat{u}_{j}) \cos{2\xi} \, ,
\end{array}
\end{equation}
which in turn enables us to Fourier transform the equations for $\Delta_Q$ and $\Delta_U$:
\begin{equation}\label{eq:pol}
\begin{array}{c}
\Delta_Q(\hat{n}) = \displaystyle\int d\eta\, e^{-\tau}\dot{\tau}\, \Pi_{ij}
(\hat{\theta}_i\hat{\theta}_j - \hat{\varphi}_i\hat{\varphi}_j)\, ,\\
\Delta_U(\hat{n}) = \displaystyle\int d\eta\, e^{-\tau}\dot{\tau}\,
\Pi_{ij} (\hat{\theta}_i\hat{\varphi}_j +
\hat{\varphi}_i\hat{\theta}_j)\, .
\end{array}\end{equation}
The above equations for the polarization along the line of
sight (\ref{eq:pol}), together with the modified temperature
equation (\ref{eq:temp}) are the desired results.
Of course, the equations do not contain new physics, because, as
mentioned earlier, it is included
in Einstein-Boltzmann solvers that
directly compute CMB power spectra such as
\textsc{cmbfast}~\cite{seljak96} but, unlike harmonic space formulas,
their form gives us a clear view
of the phenomena producing CMB temperature anisotropies and
polarization along the line of sight.
Given a 3D Einstein-Boltzmann solver to evolve the cosmological
perturbations such as the Landriau-Shellard code \cite{methods}, they also allow us to
compute CMB maps from the recombination epoch onwards, by ray tracing
through the simulation box.
In particular, they can be used in the case where they are seeded by
cosmic strings~\cite{hrcmb,pol}, which allows one to search for
stringlike features in the CMB maps.

\section{Conclusion}

I have derived real space formulas for the CMB temperature
anisotropies and polarization that contain all the relevant physics to
first order in the cosmological perturbations.  These are equivalent
to the harmonic space formulas found in the literature and
furthermore, allow the direct
computation of maps of the CMB polarization and temperature from the
recombination epoch onwards when the perturbations are evolved in
three-dimensional simulations.  This allows to study localized
features in the maps that are not identifiable in the power spectra.

\section*{ACKNOWLEDGMENTS}

I would like to thank Eiichiro Komatsu for suggesting this
topic of investigation many years ago
and showing me the way
by deriving the scalar part of Eq. (\ref{eq:temp}).

\end{document}